\documentclass[twocolumn,showpacs,aps,floatfix]{revtex4}

\usepackage{graphicx}
\usepackage{bm}
\usepackage{amsmath}
\usepackage{color,colordvi} 
\newcommand{\ignore}[1]{}

\def\beq{\begin{equation}}
\def\eeq{\end{equation}}
\def\beqa{\begin{eqnarray}}
\def\eeqa{\end{eqnarray}}

\def \tr{\hbox{Tr}}

\newtheorem{theorem}{Theorem}

\begin{document}

\title{Quantum and thermal fluctuations in bosonic Josephson junctions}

\author{B. Juli\'a-D\'\i az$^{1,2}$, A. D. Gottlieb $^3$ J. Martorell$^1$, 
and A. Polls$^1$}

\affiliation{$^1$Departament d'Estructura i Constituents de 
la Mat\`eria, Facultat de F\'isica, U. Barcelona, 08028 
Barcelona, Spain}
\affiliation{$^2$ICFO-Institut de Ci\`encies Fot\`oniques,
Parc Mediterrani de la Tecnologia, 08860 Barcelona, Spain}
\affiliation{$^3$Wolfgang Pauli Institute, Nordbergstrasse 
15, 1090 Vienna, Austria}

\begin{abstract}

We use the Bose-Hubbard Hamiltonian to study quantum fluctuations 
in canonical equilibrium ensembles of bosonic Josephson junctions at 
relatively high temperatures, comparing the results for finite particle numbers to the classical limit that is attained as $N$ approaches 
infinity.  We consider both attractive and repulsive atom-atom 
interactions, with especial focus on the behavior near the $T=0$ quantum 
phase transition that occurs, for large enough $N$, when attractive 
interactions surpass a critical level.  Differences between Bose-Hubbard 
results for small $N$ and those of the classical limit are quite small 
even when $N \sim 100$, with deviations from the limit diminishing as $1/N$. 
\end{abstract}

\date{\today}

\maketitle

\section{Introduction}

Bosonic Josephson junctions (BJJ) provide a versatile setup 
for exploring correlated quantum many-body states, such as 
pseudo-spin squeezed states and Schr\"odinger's cat-like 
states~\cite{Mil97,Smerzi97,cirac98,jav99,raghavan99,LE01,jame05,mueller06,ours10,mom10}. 
Moreover, the relations between fully quantal, semiclassical and 
classical models of BJJ provide insights into phenomena such as 
dynamical quantum tunneling or quantum chaos~\cite{grae08,vardi13}.

Schematically, a bosonic Josephson junction consists of an 
ultracold atomic cloud in which (i) idealized atoms can populate 
only two single-particle modes, or levels (ii) atoms can hop independently 
from one level to the other, and (iii) atoms interact 
with each other only locally (through contact-like atom-atom 
interactions). The main differences between existing experimental 
realizations stem from the nature of the two levels. 
In ``external" Josephson junctions, the two levels are spatially 
separated modes~\cite{albiez05}, whereas in ``internal'' Josephson 
junctions, the levels are spin degrees of freedom internal to the  
atoms~\cite{zib10}; see the recent Ref.~\cite{gross12} for a 
comprehensive tutorial. To a good approximation, the many-body 
Hamiltonian describing a BJJ can be written as a two-site Bose-Hubbard 
(BH) Hamiltonian~\cite{Mil97}: 
\begin{eqnarray}
H_{BH} &=& 
-J (\hat{a}_1^{\dag} \hat{a}_2 
             + \hat{a}_2^{\dag} \hat{a}_1)\nonumber \\
&+& 
\frac{U}{2}(\hat{a}_1^{\dag} \hat{a}_1^{\dag} \hat{a}_1 \hat{a}_1 
+ \hat{a}_2^{\dag} \hat{a}_2^{\dag} \hat{a}_2 \hat{a}_2 ) \ .
\label{eq:bh1}
\end{eqnarray} 
with $[\hat{a}_i, \hat{a}_j^{\dag}] = \delta_{i\ j}$. The first term 
models ``hopping'' between levels $1$ and $2$, with strength given 
by the linear coupling energy $J$. The second term accounts for the 
interaction between the atoms. This many-body Hamiltonian can also 
be regarded as a particular case of the Lipkin-Meshkov-Glick 
model~\cite{lipkin}. Experiments confirm the ability of this 
Hamiltonian to describe the ground state of BJJ and to predict 
dynamics~\cite{albiez05,esteve08,gross10,zib10,zibphd}.

In these experiments one exercises some control over the three main 
parameters of $H_{BH}$: the atom number $N$, the linear coupling $J$ 
and the atom-atom interaction strength $U$.  The strength of the 
atom-atom interaction is measured by the dimensionless parameter 
\beq
 \gamma  \ =\  \frac{N U }{ 2 J}\ .
\eeq
There is an interesting quantum phase transition at $\gamma = -1$, 
beyond which the attractive atom-atom interactions cause a bifurcation 
in the ground state properties of the system in the semiclassical 
limit~\cite{cirac98,ours10,ours11}. State-of-the-art experiments can deal 
with $N$ down to $\sim$ 300~\cite{zibphd}, with $N J$ varying by 
several orders of magnitude~\cite{esteve08} and $U$ varying over 
a wide range~\cite{zib10} including both attractive and repulsive 
atom-atom interactions. This vast freedom allows one, in principle, 
to study the stable formation of ``cat'' states under attractive 
interactions~\cite{cirac98,ours10,ours11} and highly squeezed spin 
states under repulsive interactions~\cite{kita}.

To create such many-body states in a laboratory BJJ, especially external 
BJJs, one must manage the effects of temperatures 
$T>0$ ~\cite{albiez05,gross12}. For example, temperature effects 
still present experimental challenges 
against production of highly spin-squeezed states~\cite{gross12}. At 
finite temperature $T>0$ the appropriate state to study is a canonical 
equilibrium ensemble in which a large number of eigenstates of the many-body 
system are significantly populated. The effects of temperature on the 
coherence of the Josephson junction have been studied both theoretically 
and experimentally in Ref.~\cite{PS01,GO07,GS09}. Here, we shall assume 
values of $T$ that are comparable to the total energy in the BJJ, but low 
enough that the system remains bimodal to a good approximation (a condition 
which may or may not be fulfilled, depending on the actual implementation 
of the BJJ).

We have studied the effects of temperature in BJJs by numerical 
diagonalization of $H_{BH}$ for $N \sim 100$. When $(NJ)/(k_B T) \ll N $, 
the effect of temperature can be approximated by the classical 
($N \to \infty$) theory of Gottlieb and Schumm~\cite{GS09}, 
provided also that $|\gamma| \ll N$. We have found that the 
Gottlieb-Schumm (GS) predictions are remarkably close to the ``exact" 
BH results both for attractive and repulsive interactions, even 
when $N$ is only $100$ or less.  

The GS theorem asserts that the finite 
temperature equilibrium ensembles of the two-mode model $H_{BH}$ 
ressemble classical statistical mixtures of coherent quantum 
states~\cite{GS09}. Coherent quantum states of $N$ two-mode bosons 
are in one-to-one correspondence with points on the Bloch sphere, 
see Eq.~(\ref{eq:cohs}) below; coherent states can be mixed or 
averaged according to any measure on the Bloch sphere. As the 
number $N$ of bosons increases, the kinetic energy of a coherent 
state scales as $N$ and its potential energy scales as $N^2$. In 
a limit where $NJ/(k_BT)$ and $U N^2/(2 k_BT)$ converge to finite 
limits $\delta$ and $\epsilon$ as $N \to \infty$, the canonical 
thermal $N$-boson ensembles converge towards the mixture of coherent 
states that has the following (unnormalized) density function on 
the Bloch sphere:
\begin{eqnarray}
{\cal P}(\theta, \phi )
&\propto& {\rm  exp} (\delta \sin \theta \cos \phi - \varepsilon
\cos^2 \theta/2)\,.
\label{unnormalized}
\end{eqnarray}
As shown in the appendix, this density function is the limit of the 
normalized Husimi distributions of the finite temperature equilibrium ensembles.  

In Ref.~\cite{GS09} the convergence of the finite temperature 
equilibrium ensembles to the mixture of coherent states was proved 
only in the following, rather limited, sense: that the expected  
values of $k$-particle observables, with $k$ remaining finite as 
$N \to \infty$, converge to the expectations with respect to the 
mixture~(\ref{unnormalized}) of coherent states. This kind of convergence 
is too weak to distinguish, for example, the entangled ``NOON" state 
$ \tfrac{1}{\sqrt{2}}|N,0\rangle + \tfrac{1}{\sqrt{2}}| 0,N\rangle$
from the mixed density  
$\tfrac12 |N,0\rangle\!\langle N,0| + \tfrac12 |0,N\rangle\!\langle 0,N|$, 
for both of them have the same $k$-particle correlations if $k < N$.    

As the GS theory is an essentially ``classical'' theory concerning an 
$N\rightarrow \infty$ limit, it does not describe the quantum fluctuations 
due to finite $N$.  The finite-$N$ corrections to the classical theory 
are generally expected to be proportional to $1/N$ (see 
formula~(\ref{eq:po2}) below). 
The differences we observe between the Bose-Hubbard solutions and the 
classical predictions do appear to diminish as $1/N$ in a controlled 
fashion, i.e., without large coefficients, even for $N < 100$.     

The predictions of GS theory break down at temperatures of the 
order of the hopping energy, when only a few of the lowest 
eigenstates are sizeably occupied. For such low temperatures, and 
for strong repulsive atom-atom interactions, one may take a Bogoliubov 
theory approach as in Ref.~\cite{ober06}, or use the $1/N$ model 
studied in Ref.~\cite{malo13}, which yields similar results.

The rest of this article is organized as follows: First, in 
Sec.~\ref{sec:model} we review the two-site BH Hamiltonian, 
coherent states of systems of two-mode bosons, and Husimi 
distributions of their states. In Sec.~\ref{temperature section} 
we present our data showing effects of finite temperature. 
In Sec.~\ref{sec:cm} we review the classical theory and compare 
its predictions to the exact BH results. Our conclusions are stated 
in Sec.~\ref{sec:sum} and a proof of the GS theorem is given in the appendix.

\section{The two-site Bose-Hubbard Hamiltonian} 
\label{sec:model}

\begin{figure} [t]                         
\vspace{10pt}
\includegraphics[width=7cm,angle=-0]{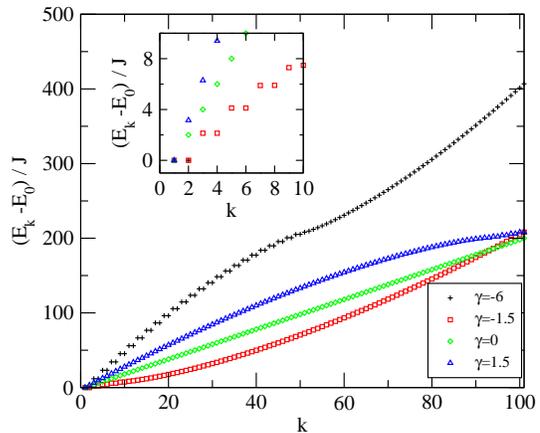}
\caption{
(Color online) Energy spectra for $N=100$ atoms and several 
values of $\gamma=-6,-1.5, 0, 1.5$. Only the excitation 
energies are plotted. The inset focuses in the 
lower eigenstates.
}
\label{fig:spec}
\end{figure} 

\subsection{Spectral properties of the Bose-Hubbard Hamiltonian} 
\label{spectral properties}

The eigenfunctions and eigenvalues of the BH Hamiltonian (\ref{eq:bh1}) 
have been studied intensively~\cite{Mil97,cirac98,jame05,vardi,ST08,ours10}. 
For our subsequent discussion we only need to point out a couple of their 
properties. Figure~\ref{fig:spec} displays eigenvalues of $H_{BH}$ for 
several choices of $\gamma$. Note that, except for the appearance of 
quasi-doublets, the energy levels increase smoothly. Also, the figure 
exhibits, for $\gamma=\pm 1.5$, the symmetry in the spectral properties 
between repulsive (r) and attractive (a) interactions, 
$E_k^{(\rm r)}-E_0^{(\rm r)} = E_N^{(\rm a)} -E_{N-k}^{(\rm a)}$. The latter 
can be seen by noting that the BH Hamiltonians for attractive and 
repulsive interactions are related by a rotation around $\hat{J}_y$ 
of angle $\pi$ and an overall sign. The eigenstates for the repulsive 
and attractive case are also easily related by the same rotation.

\subsection{Pseudo-spin formalism and coherent states} 

The state of a BJJ can be described by a large spin subject to  
a Hamiltonian that has both a linear and a non-linear term. 
As is customary~\cite{LE01}, we introduce the ``pseudo-spin'' operators
\begin{eqnarray}
{\hat J}_x &=& \frac{1}{2}( \hat{a}_1^{\dag} \hat{a}_2 
+ \hat{a}_2^{\dag} \hat{a}_1) \nonumber \\
{\hat J}_y &=& \frac{1}{2i}( \hat{a}_1^{\dag} \hat{a}_2 
- \hat{a}_2^{\dag} \hat{a}_1) \nonumber \\
{\hat J}_z &=& \frac{1}{2}( \hat{a}_1^{\dag} \hat{a}_1 
- \hat{a}_2^{\dag} \hat{a}_2) 
\label{eq:bh1b}
\end{eqnarray}
which satisfy the angular momentum commutation relations and 
\[
    {\hat J}_x^2 + {\hat J}_y^2 + {\hat J}_z^2 = (N^2+2N)/4
\]
on the $N$-boson space.
In terms of these pseudo-spin operators, 
\begin{eqnarray}
H_{BH} 
&=& - 2 J {\hat J}_x + U  {\hat J}_z^2 + U
\big(\tfrac{N^2}{4}-\tfrac{N}{2}\big) \ .
\label{eq:bh3}
\end{eqnarray}

For any unit vector ${\bf u} = (u_x,u_y,u_z)$, let 
$$ {\hat J}_{\bf u}= u_x{\hat J}_x + u_y{\hat J}_y + u_z{\hat J}_z\ .$$  
The eigenvectors of the pseudo-spin operators ${\hat J}_{\bf u}$ are the 
``coherent states'' wherein all $N$ particles occupy the same mode~\cite{husiref}. 
Specifically, if $(\theta,\phi)$ are the spherical coordinates of a 
unit vector ${\bf u}$, so that 
\begin{eqnarray*}
  u_x & = & \sin\theta \cos\phi \\
  u_y & = & \sin\theta \sin \phi \\
  u_z & = & \cos\theta \ ,
\end{eqnarray*}
then the coherent state $| \Psi_{\theta,\phi}^N \rangle$ defined by 
\beq
 | \Psi_{\theta,\phi}^N \rangle  = 
 {1\over \sqrt{N!}} \left(\cos \theta/2 \ a^{\dag}_1 
+ e^{i  \phi} \sin \theta/2 \ a^{\dag}_2 \right)^N |\O \rangle 
\label{eq:cohs} 
\eeq
is an eigenvector of ${\hat J}_{\bf u}$ with eigenvalue $N/2$.

Any pure state of $N$-bosons in a two-mode BJJ can be written as superposition 
of coherent states, for example by using the completeness 
relation~\cite{husiref}
\begin{equation}
\frac{1}{4\pi}\int_0^\pi  \sin\theta d\theta \int_0^{2\pi} d\phi 
\ |\Psi_{\theta,\phi}^N \rangle\!\langle\Psi_{\theta,\phi}^N |  \ = \frac{1}{N+1}
\label{completeness}
\end{equation}
(note that $\sin\theta d\theta d\phi$ is the surface area element on 
the unit sphere). Though the coherent states form a complete set of 
$N$-particle states, they do not constitute an orthonormal basis; two 
coherent states are not orthogonal unless they correspond to antipodal 
points on the sphere.

We will also use the notation 
$(\hat{x},\hat{y},\hat{z})=\frac{2}{N} (\hat{J}_x,\hat{J}_y,\hat{J}_z)$ 
for normalized pseudo-spin operators. The observable $\hat{z}$ is the 
population imbalance operator between the modes $\hat{a}_1^{\dag}|\O \rangle$ 
and $\hat{a}_2^{\dag}|\O \rangle$, normalized to have values 
between $-1$ and $+1$. The angular momentum commutation relations imply that  
\begin{equation}
 [\hat{x},\hat{y}] = 
\tfrac{2}{N}i\hat{z}, \  \hbox{etc.,} \quad {\rm and}\;
\hat{x}^2+ \hat{y}^2 +\hat{z}^2 = 1+\tfrac{2}{N}\ ,
\label{eq:po2}
\end{equation}
which indicates that the average spin projections behave like 
classical observables in the limit $N \rightarrow \infty$.

\subsection{Husimi distributions}

\begin{figure} [t]                         
\vspace{10pt}
\includegraphics[width=9.5cm,angle=-90]{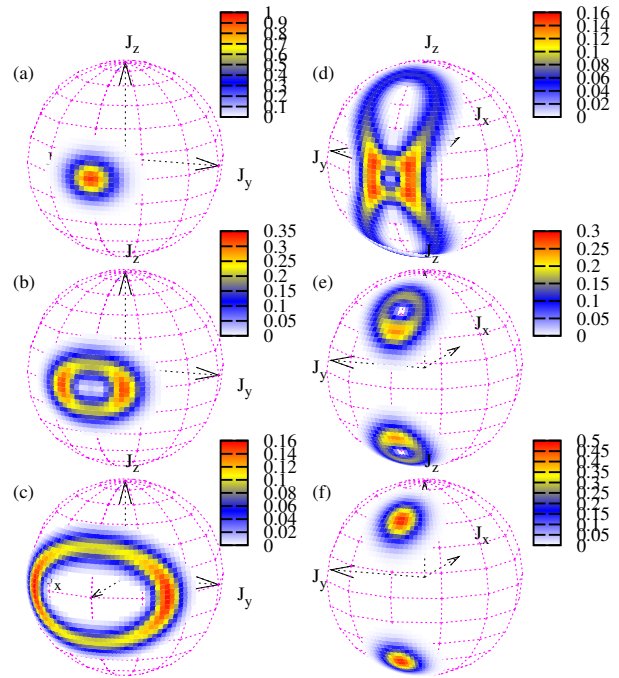}
\caption{
(Color online) Husimi distributions of some eigenvectors of the BH for 
$\gamma=1.5$. In panels (a), (b), (c), (d), (e), and (f), we plot the 
distributions corresponding to the states with index $k=0, 2, 10, 90, 98$, 
and 101, where $k=0$ is the ground state. $N=100$ particles. 
}
\label{fig:hueva}
\end{figure}

When a state of a system of $N$ two-mode bosons is 
represented by a density matrix ${\hat\rho}$, the Husimi distribution, or Q representation,  
of that state is the function 
\beq
P(\theta,\phi)=  \langle \Psi_{\theta,\phi}^N | {\hat\rho} 
|\Psi_{\theta,\phi}^N \rangle 
\eeq
defined on the unit sphere \cite{lee84,mahmud05}.  

The Husimi distribution of a pure state $| \Phi \rangle$ is just 
$\big| \langle \Psi_{\theta,\phi}^N|\Phi \rangle \big|^2$. For example, the Husimi distribution of the 
coherent state $|\Psi_{0,0}^N\rangle$ is 
$\big| \langle \Psi_{\theta,\phi}^N| \Psi_{0,0}^N \rangle \big|^2 = \big(\cos^2(\theta/2)\big)^N$. 
The Husimi distribution of a coherent state 
$|\Psi_{\theta,\phi}^N\rangle$ is spread over the whole Bloch sphere, 
but as $N\rightarrow\infty$ the distribution becomes more and more 
concentrated about the point $(\theta,\phi)$. 

Husimi distributions can help one visualize eigenstates of a Hamiltonian and see their 
connection to the classical orbits.  
Fig.~\ref{fig:hueva} portrays the Husimi distribution of some eigenvectors 
of $H_{BH}$ for a characteristic repulsive interaction: $\gamma = 1.5$. 
The ground state Husimi distribution in panel (a) is similar to that 
of the coherent state corresponding to the point $(\pi/2,0)$ on the 
Bloch sphere, but is somewhat flattened, reflecting the number-squeezing due to the repulsive interaction, as studied, e.g. in Ref.~\cite{oursober}. As the energy 
$E_i$ increases, the first Husimi distributions, graphed in panels (b) 
and (c), have the shape of elliptical rings of increasing size. Panels 
(e) and (f) show that the highest energy states have their Husimi 
distributions concentrated around the other classical stationary 
points of the Hamiltonian~\cite{raghavan99}. Panel (d) shows the Husimi 
distribution before the transition to the highest energy states. 

In this article, we are concerned with 
thermal equilibrium ensembles for $H_{BH}$, whose density matrices are 
the canonical ones. Thus, the Husimi distribution for the canonical 
ensemble of $N$ bosons at temperature $T$ is 
\beq
P_{N,T}(\theta,\phi) \ = \  
{\cal Z}^{-1} \ \sum_{i=0}^N e^{-E_i/k_BT} 
\big| \langle \Psi_{\theta,\phi}^N|\Phi_i \rangle \big|^2
\label{Husimi of canonical}
\eeq
where $|\Phi_i\rangle$ is the $i^{th}$ eigenstate of $H_{BH}$, 
with energy $E_i$, and
\[
{\cal Z}=\sum_{i=0}^N e^{-E_i/k_BT} 
\]
is the partition function. 
Note that the functions $\big| \langle \Psi_{\theta,\phi}^N|\Phi_i \rangle \big|^2$ appearing in (\ref{Husimi of canonical}) are the Husimi distributions of the eigenstates.

\section{Temperature effects}
\label{temperature section}

\begin{figure} [t]                         
\vspace{10pt}
\includegraphics[width=9.5cm,angle=-90]{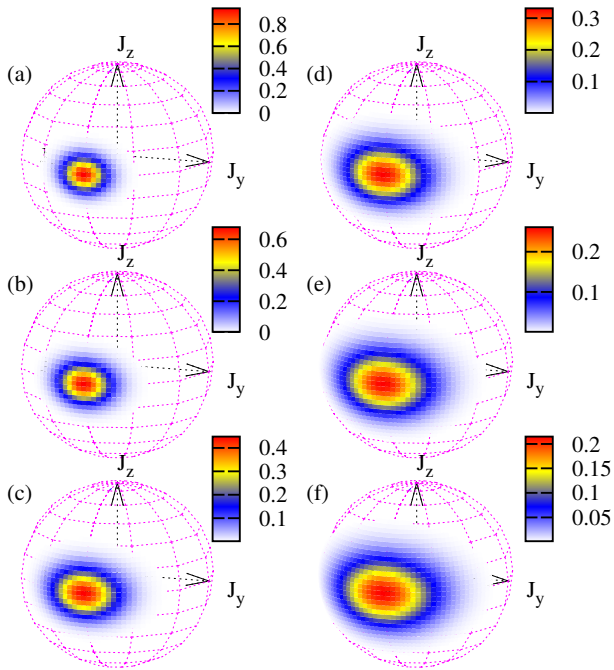}
\caption{(Color online) Husimi distributions for $\gamma=1.5$ 
(repulsive interactions). From (a) to (f) the temperatures are 
$k_BT/(NJ)=0, 0.25, 0.5, 0.75, 1., 1.25$ and 1.5. $N=100$. }
\label{fig:hu2-20r}
\end{figure}

\begin{figure} [t]                         
\vspace{10pt}
\includegraphics[width=9.5cm,angle=-90]{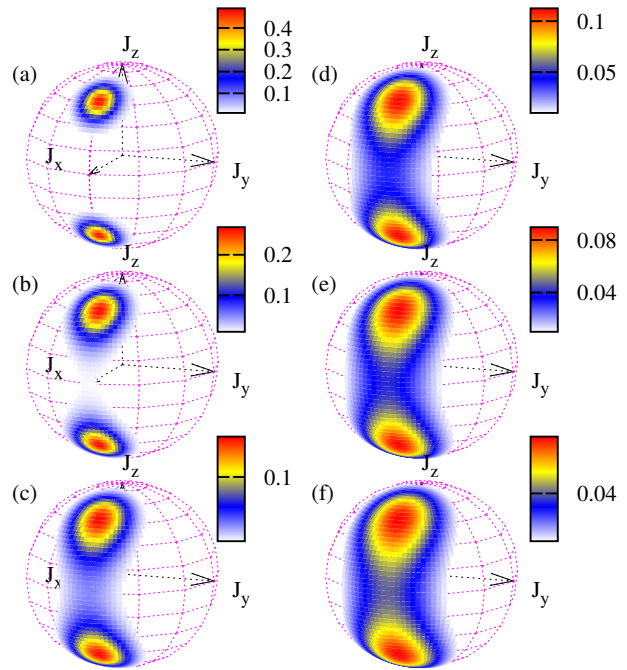}
\caption{(Color online) Husimi distributions for $\gamma=-1.5$ 
(attractive interactions). From top to bottom and left to right 
the temperatures are $k_BT/(NJ)=0, 0.25, 0.5, 0.75, 1., 1.25$ 
and 1.5. $N=100$. }
\label{fig:hu2-20a}
\end{figure}

In this section we focus on the high temperature equilibrium behavior 
of BJJs in the Rabi-Josephson boundary regime~\cite{GS09} where 
$|\gamma| \sim 1$. This regime can now be addressed experimentally 
in internal BJJs, where Feshbach resonances can be used to tune 
the interaction strength, and where the number of atoms is also 
relatively small (in the hundreds)~\cite{zib10}. In the experiments 
reported in~\cite{zibphd}, for example, the number of atoms was around 
300 and it was possible to control the value of $\gamma$ about 
$\gamma = -1$. We focus on the same regime of small $\gamma$ and relatively 
small $N$, but we consider relatively high temperatures $k_BT \sim NJ$. 
These temperatures are much higher than those relevant to the recent 
experiments just mentioned, where $k_BT \sim J$.  Nevertheless, our 
exploration of the higher temperature behavior of the two-mode model 
should provide guidance for future experiments that may be performed 
in this regime.

To take a first look at the effect of higher temperatures in BJJs, 
we display the Husimi distributions of some of the canonical 
thermal equilibrium ensembles.   

Figures~\ref{fig:hu2-20r}  and~\ref{fig:hu2-20a}  illustrate the change of the 
Husimi distributions of the canonical equilibrium ensembles as $T$ increases 
moderately. When $\gamma = +1.5$ one sees that the roughly elliptical shape 
remains, but covers a greater area on the Bloch sphere. The Husimi 
distributions for attractive interactions, shown in Fig.~\ref{fig:hu2-20a} for $\gamma=-1.5$, 
resemble the shapes of the Husimi distributions of the higher energy 
eigenstates for {\it repulsive} interactions, shown in panels 
(d)-(f) of Fig.~\ref{fig:hueva}. The reason for this is that the 
Husimi distribution of a thermal equilibrium distribution is a 
mixture of the Husimi distributions of the lower energy eigenstates 
(cf., formula (\ref{Husimi of canonical})), and the Husimi distributions of low 
energy eigenstates for attractive interactions are identical to those of the 
corresponding high energy eigenstates for repulsive interactions, due to the 
spectral properties of the two-site Bose-Hubbard Hamiltonian mentioned 
in Sec.~\ref{spectral properties}.

We turn now to look at the behavior of the observables ${\hat x}$, ${\hat y}$, 
and ${\hat z}$.  The average values $\langle {\hat x}\rangle$ and 
$\langle {\hat z}^2 \rangle$ are of especial interest~\cite{GO07}. 
They are used for quantifying spin-squeezing~\cite{sore01,esteve08,gross12} 
and for noise thermometry~\cite{ober06,noise_thermometer06,GS09} in BJJs.    

\begin{figure} [tb]                         
\includegraphics[width=8.cm]{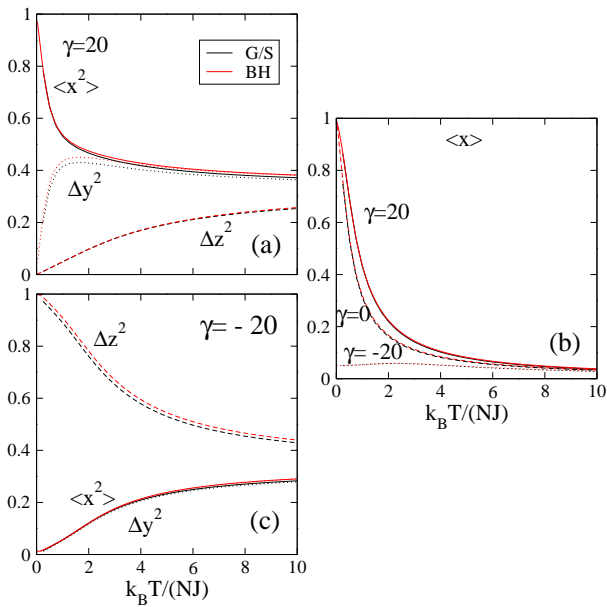}
\caption{
(Color online) Thermal averages for 
$\gamma=-20, 0$ and $20$. Black lines show the prediction 
with the classical $N\to \infty$ approximation of Ref.~\cite{GS09} 
while red ones are the BH results with $N=100$.}
\label{fig:bhgs-repulsive}
\end{figure}

\begin{figure} [tb]                         
\includegraphics[width=8.cm]{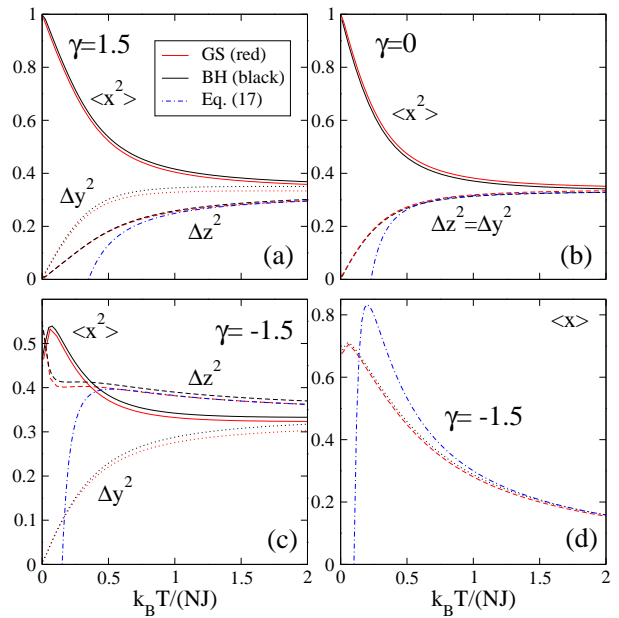}
\caption{
(Color online) Thermal averages for $\gamma=-1.5, 0$ and $1.5$. 
Red lines are the prediction with the classical $N\to \infty$ 
approximation of Ref.~\cite{GS09} while black ones are the BH results with $N=100$. The blue dot-dashed lines 
are the asymptotic GS predictions for large temperature given 
in~(\ref{averagex}) and~(\ref{averagez^2}).}
\label{fig:bhgs-attractive}
\end{figure}

The average $\alpha \equiv \langle \hat{x}\rangle$, called the coherence 
factor~\cite{PS01,GO07}, is proportional to the mean fringe visibility in 
interference experiments. The coherence factor $\alpha$ is plotted in panels  
(b) and (d) of Figs.~\ref{fig:bhgs-repulsive} and~\ref{fig:bhgs-attractive}, 
respectively. We will consider the lower temperature ($k_BT \sim J$) 
behavior of the $\alpha$ in Sec.~\ref{sec:cm} below. 

The averages $\langle \hat{y}\rangle$ and $\langle \hat{z}\rangle$ are both 
equal to $0$ in the absence of any bias affecting ${\hat J}_y$ and 
${\hat J}_z$. For this reason, $\langle \hat{y}^2 \rangle$ and 
$\langle \hat{z}^2 \rangle$ quantify the fluctuations of the observables ${\hat y}$ 
and ${\hat z}$ about their averages, and we shall accordingly use the notation 
$\Delta {\hat y}^2$ and $\Delta {\hat z}^2$ instead of $\langle \hat{y}^2 \rangle$ 
and $\langle \hat{z}^2 \rangle$. However, note that 
$\Delta {\hat x}^2 = \langle {\hat x}^2 \rangle - \langle {\hat x} \rangle^2 $ 
is not the same as $\langle {\hat x}^2 \rangle$ because 
$\langle {\hat x} \rangle \neq 0$.

\begin{figure} [tb]                         
\includegraphics[width=8.cm, angle=-90]{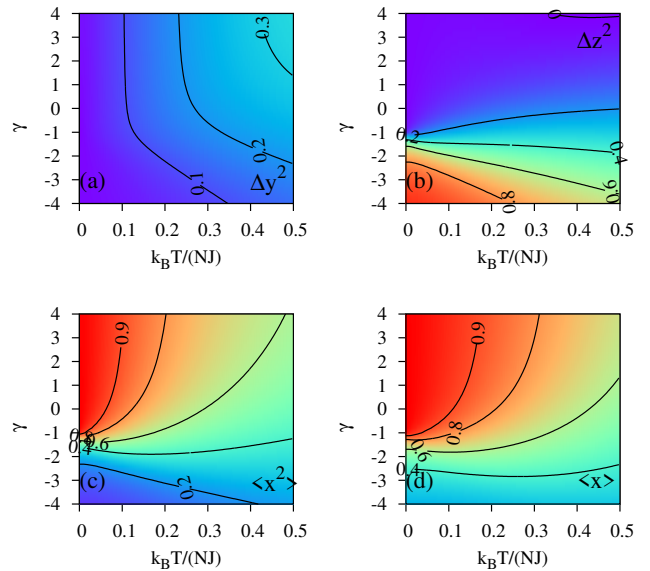}
\caption{
(Color online) Contour plots of 
$\Delta \hat{y}^2$,  
$\Delta\hat{z}^2 $, 
$\langle \hat{x}^2\rangle$, and $\langle \hat{x}\rangle$ computed 
for Bose-Hubbard model with $N=100$, showing dependence on $T$ and $\gamma$. 
}
\label{fig:contour}
\end{figure} 

To provide the overall picture, Fig.~\ref{fig:contour} displays the dependence 
of $\langle \hat{x} \rangle$, $\langle \hat{x}^2 \rangle$, $\Delta \hat{y}^2$, 
and $\Delta \hat{z}^2$ on both $\gamma$ and $k_BT /(NJ)$. The figure is 
made for a fixed value of $N$, varying $\gamma$ and $T$. Note the 
abrupt change of behavior around $\gamma=-1$ and $T=0$. This reflects 
the bifurcation in the ground state properties there~\cite{zib10,ours10}. 
For values of $\gamma<-1$ we have that $\langle \hat{x}^2\rangle$ 
and $\Delta \hat{y}^2$ remain small as $T$ increases, while 
$\Delta \hat{z}^2$ decreases from a value close to $1$ at $T=0$.  On the 
other hand, for values of $\gamma>-1$, it is now $\langle \hat{x}^2\rangle$ 
that is close to 1 near $T=0$, while $\Delta \hat{y}^2$ and 
$\Delta \hat{z}^2$ are small.

Panels (a) and (c) of Figs.~\ref{fig:bhgs-repulsive} 
and~(a), (b) and (c) of~\ref{fig:bhgs-attractive} are plots of 
$\langle \hat{x}^2 \rangle$, $\Delta \hat{y}^2$, and $\Delta \hat{z}^2$ for 
$\gamma =0, \pm 1.5$, and $\pm 20$. Note how quickly $\Delta {\hat z}^2$ 
drops with increasing $T$ when $\gamma =-1.5$ (panel (c) of 
Fig.~\ref{fig:bhgs-attractive}). At zero temperature, $\Delta \hat{z}^2 >0$ 
because the ground state is cat-like, but with a small increase in 
temperature $\Delta {\hat z}^2$ quickly drops 20\% (with a concomitant 
increase in $\langle x^2\rangle$ and $\Delta {\hat y}^2$ due to the 
relation~(\ref{eq:po2})).

\section{Comparison with classical theory}
\label{sec:cm}

In this section, we review the classical, i.e., $N\rightarrow \infty$, limit 
for the regime $k_BT \sim NJ$ and $|\gamma| \sim 1$. 
We shall see that the predictions of the classical theory are remarkably 
accurate even when $N$ is rather small.

\subsection{The classical theory}
\label{review of GS theory}

Following~\cite{GS09}, we define dimensionless ratios which measure the 
tunneling and interaction energies with respect to 
the thermal energy, $k_B T$: 
$$\delta= \frac{NJ}{k_BT}   \qquad \hbox{and} 
\qquad \varepsilon = \frac{ N^2U}{2 k_BT}\ .$$
Note that $\gamma = \epsilon/\delta$.
We prove in the appendix that the normalized Husimi distributions of the canonical 
equilibrium ensembles converge to a function proportional to 
\begin{eqnarray}
Q_{\delta,\epsilon}(\theta,\phi )  & = & \exp(\delta \sin\theta\cos\phi 
- \varepsilon \cos^2\theta /2) 
\nonumber \\
& = & \exp(\delta x - \varepsilon z^2/2) 
\label{eq:po1}
\end{eqnarray}
in the limit $N \rightarrow \infty$ with $\delta$ and $\epsilon$ remaining 
constant.

According to the theorem stated in~\cite{GS09}, canonical thermal 
ensemble averages of certain observables also converge in this limit. 
Let $O(\hat{x},\hat{y},\hat{z})$ denote any polynomial in the 
operators $\hat{x}$, $\hat{y}$, and $\hat{z}$, and let 
$\langle O(\hat{x},\hat{y},\hat{z})\rangle_{N,T}$ denote the 
canonical average value of the corresponding observable. Then 
$ \langle O(\hat{x},\hat{y},\hat{z})\rangle_{N,T} $ tends to 
\beq
 \frac{1}{\cal N} \int_S e^{\delta x - \varepsilon z^2/2} \;  O(x,y,z)  \ dS
 \label{eq:ave}
\eeq
as $N \rightarrow \infty$ while $\delta$ and $\epsilon$ remain constant, 
where $dS$ denotes surface area measure on the unit sphere
$ 
   S \ = \ \{ (x,y,z): \ x^2 + y^2 + z^2 = 1 \}
$ 
and 
\beq
 {\cal N} \ = \   \int_S e^{\delta x - \varepsilon z^2/2} dS\ .
\label{normalizing const}
\eeq

The preceding theorem is also proved in the appendix, in a somewhat 
more general form. For the present, {\it assuming} that canonical 
equilibrium ensembles behave like statistical mixtures of coherent 
states when $N$ is large, let us explain where the weight function 
$\exp(\delta x - \varepsilon z^2/2) $ comes from:     

In terms of the operators ${\hat x}, {\hat y}$ and ${\hat z}$, 
the Hamiltonian reads  
\[ 
H_{BH} \ = \ -N J {\hat x} + \tfrac{U}{4} N^2 {\hat z}^2 
+ U \big(\tfrac{N^2}{4}-\tfrac{N}{2}\big) \ .
\]
The preceding Hamiltonian operator is considered as acting only 
upon the $N$-particle subspace of the boson Fock space. Accordingly, 
we drop the constant term and introduce $N$ explicitly into the 
notation for the Hamiltonian, defining 
\[
H_N 
\ = \  -N J {\hat x} + \tfrac{U}{4} N^2 {\hat z}^2 \,.
\]
In thermal equilibrium at temperature $T$, the statistical weight 
of the coherent state centered at the point $(\theta, \phi)$ in 
the Bloch sphere should be proportional to 
$ \exp\big(-\tfrac{1}{k_BT}\langle \Psi_{\theta,\phi}^N | 
H_N | \Psi_{\theta,\phi}^N \rangle \big) $ because 
$\langle \Psi_{\theta,\phi}^N | H_N | \Psi_{\theta,\phi}^N \rangle$ is the 
energy of the coherent state. This equals 
\beq
\exp\big( {\langle \Psi_{\theta,\phi}^N | \delta {\hat x} 
- \varepsilon{\hat z}^2/2 | \Psi_{\theta,\phi}^N \rangle} \big)
\label{labelme}
\eeq
since 
$- \tfrac{1}{k_BT}H_N  = \delta \ {\hat x} - \varepsilon\ {\hat z}^2/2 $. 
In the classical limit,~(\ref{labelme}) becomes 
$\exp(\delta x - \varepsilon z^2/2)$.

Using~(\ref{eq:ave}) one can compute expectations and variances of 
observables of interest. For example, one can calculate the coherence 
factor as follows. Defining $a \equiv \sqrt{1-x^2}$, the integral 
in~(\ref{normalizing const}) is
\[
{\cal N} 
\ = \  2 \int_{-1}^1 \ e^{\delta x}  \int_{-a}^a  \ 
\frac{e^{-\varepsilon(a^2-y^2)/2}}{\sqrt{a^2-y^2}} \ dy \ dx \ ,
\]
where the factor $2$ takes into account the equal contributions 
of points with $z>0$ and $z<0$.  Next, change $y = a \cos \xi$ so that 
$y \in (-a,a)$ implies that $\xi \in(0,\pi)$. Then,  
\[ 
{\cal N} 
\ = \  \int_{-1}^1 \  I_0\left(\tfrac{\varepsilon}{4}(1-x^2)\right) 
\ e^{\delta x +\varepsilon  x^2/4} \ dx\ .
\]
Using the same changes of variable to rewrite the integral 
$\int x e^{\delta x - \varepsilon z^2/2} $, one arrives at the formula 
for the coherence factor given in Ref.~\cite{GS09}:
\beq
\alpha_{\delta,\epsilon} \ = \  \frac{\int_{-1}^1 \ x \  I_0
\left(\frac{\varepsilon}{4}(1-x^2)\right) \ 
e^{\delta x +\varepsilon     x^2/4} \ dx} 
{\int_{-1}^1 \  I_0\left(\frac{\varepsilon}{4}(1-x^2)\right) 
\ e^{\delta x +\varepsilon  x^2/4} \ dx} \ .
\label{eq:po7}
\eeq
This formula for the coherence factor generalizes the one obtained 
in Ref.~\cite{PS01} for the Josephson regime $\gamma \gg 1$.

When $k_BT$ is large compared to the energy parameters $NJ$ and $N^2U$, 
the dimensionless parameters $\delta$ and $\epsilon$ are small, and 
expected values as in~(\ref{eq:ave}) can be approximated by polynomials 
in $\delta$ and $\epsilon$. For example, using the formulas
\beqa
     \langle {\hat x} \rangle_{\delta,\epsilon} & = & 
\frac{\partial}{\partial\delta}  
 \langle \log Q_{\delta,\epsilon} \rangle \nonumber \\
\langle {\hat z^2} \rangle_{\delta,\epsilon} & = & 
-2\frac{\partial}{\partial\epsilon}  
 \langle \log Q_{\delta,\epsilon} \rangle \nonumber 
\eeqa
and expanding the exponential $\exp(\delta x - \varepsilon z^2/2)$ in 
powers of $\delta$ and $\epsilon$, calculation shows that   
\beqa
\langle \hat{x} \rangle_{\delta,\epsilon} & = & 
\tfrac13 \delta + \tfrac{1}{45}\delta\epsilon  \label{averagex} \\
\langle \hat{z}^2 \rangle_{\delta,\epsilon} & = &  
\tfrac13 -\tfrac{2}{45}\epsilon + \tfrac{2}{945}\epsilon^2 
-  \tfrac{1}{45}\delta^2 \,, \label{averagez^2} 
\eeqa
up to terms of third order in $\delta$ and $\epsilon$.

\subsection{Comparison of BH data to classical predictions}

\begin{figure} [tb]                         
\includegraphics[width=8.cm, angle=-90]{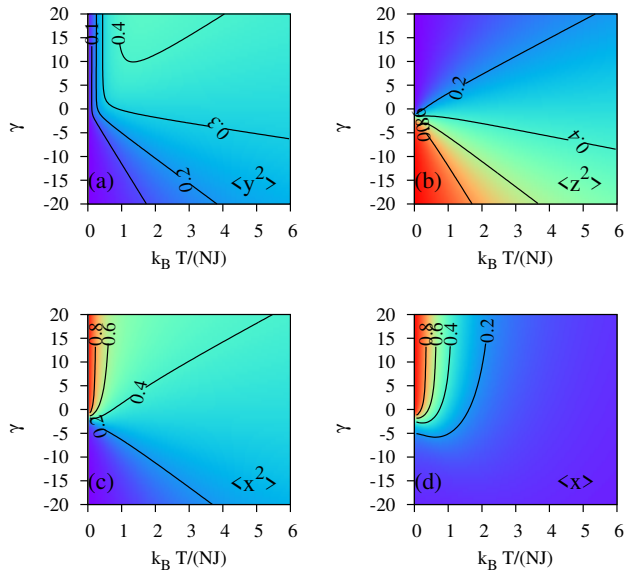}
\caption{
(Color online) Same description as Fig.~\ref{fig:contour} 
but computed with the GS approximation.
}
\label{fig:contourgs}
\end{figure} 

We now compare the results of our numerical solutions of the two-site 
Bose-Hubbard model for relatively small $N$ to the the predictions of 
the $N\rightarrow \infty$ limit discussed in the preceding paragraphs. 
In the following, ``BH'' refers to the Bose-Hubbard solutions for 
finite $N$ and ``GS" or ``classical" refers to the limit 
$N\rightarrow\infty$.

Figs.~\ref{fig:bhgs-repulsive} and \ref{fig:bhgs-attractive} show that 
the BH results are remarkably close to the classical predictions, both 
for attractive and repulsive interactions. Fig.~\ref{fig:contourgs} 
shows that the behavior around the transition at $\gamma=-1$ is well 
reproduced, as can be seen by comparing this figure with Fig.~\ref{fig:contour}. 
This figure is similar to Fig.~\ref{fig:contour} but extends 
the range of parameters to higher $T$ and $|\gamma|$.

Fig.~\ref{fig:bhgs-attractive} compares the high temperature 
GS approximations (\ref{averagex}) and (\ref{averagez^2}) to the BH results.  
These simple approximations match the BH results quite well once $k_B T > NJ$. 

\begin{figure} [tb]                         
\includegraphics[width=8.cm, angle=0]{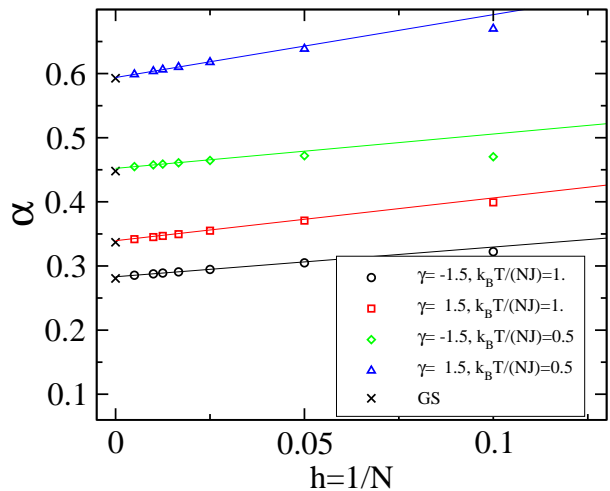}
\caption{
(Color online) Coherence factor $\alpha = \langle \hat{x} \rangle$ as 
a function of $1/N$ for two values of $k_B T/(NJ)=0.5$ and $=1$ and for two 
different values of $\gamma=\pm1.5$. The thin lines are linear regression 
fits to the first three data points.}
\label{fig:1overn}
\end{figure} 

\begin{figure} [tb]                         
\includegraphics[width=8.5cm, angle=0]{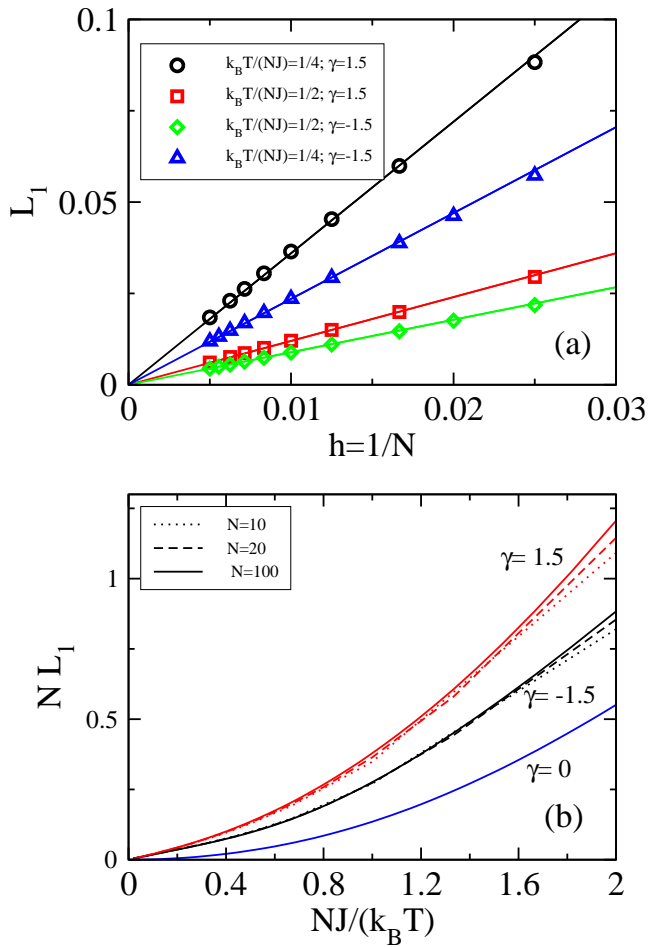}
\caption{
(Color online) (Top) Convergence of 
the BH Husimi distributions (\ref{normalized BH Husimis}) to the classical limit (\ref{normalized GS Husimi}) in terms of the averaged deviation 
(\ref{L1}). The solid lines are linear fits to the numerical results. 
(Bottom) Large temperature behavior of $N L_1$ for three 
different $N$ and $\gamma=0,\pm 1.5$. Note that we are plotting $NL_1$ 
against $\delta=NJ /k_B T$, not $J/k_BT$. The case $\gamma=0$ is computed 
with $N=100$. 
}
\label{fig:nbig}
\end{figure} 

Thus we see that the classical theory provides very good approximations even when $N \sim 100$. 
Deviations from the classical results at temperatures $k_B T\sim NJ$ 
are expected to be of order $1/N$. Fig.~\ref{fig:1overn} shows how 
coherence factor converges to the GS prediction (\ref{eq:po7}) 
as $1/N$ tends to $0$.  

Finally, we look at how the Husimi distributions (\ref{Husimi of canonical}) 
converge to their classical limit (\ref{eq:po1}). We normalize the Husimi 
distributions as is done in Ref.~\cite{lee84}, making them probability 
density functions on the unit sphere. The normalized Husimi functions
\begin{equation}
 \tilde{P}_{N,T} \ = \ \tfrac{N+1}{4\pi} P_{N,T} 
\label{normalized BH Husimis} 
\end{equation}
with $P_{N,T}$ is as in (\ref{Husimi of canonical}), converge to  
\begin{equation}
   \tilde{Q}_{\delta,\epsilon} = \tfrac{1}{\cal N} Q_{\delta,\epsilon} 
\label{normalized GS Husimi} 
\end{equation}
with $Q_{\delta,\epsilon} $ as in (\ref{eq:po1}) and ${\cal N}$ as 
in (\ref{normalizing const}). To see the rate of convergence we plot 
\begin{equation}
  L_1 \ = \  \int_S 
\ \big| \tilde{P}_{N,T} - \tilde{Q}_{\delta,\epsilon} \big|\ dS
\label{L1}
\end{equation}
against $1/N$ in the top panel Fig.~\ref{fig:nbig}. The integrated 
absolute deviation $L_1$ appears proportional to $1/N$. The bottom 
panel of the same figure shows how the convergence rate depends 
on temperature. Plotting $NL_1$ against $NJ /k_B T$ indicates that 
$L_1 \sim C(\gamma,T)/N$  with a small coefficient $C(\gamma,T)$ 
that decreases to $0$ as $T$ increases. In the non-interacting 
case $\gamma=0$, the Husimi distribution can be obtained exactly, 
\beq
P_{N,T}^{\gamma=0}(\bar\theta)={\xi-1\over \xi^{N+1}-1} \left({1+\xi +(\xi-1)\cos\bar\theta \over 2}\right)^N
\eeq
with $\xi=e^{2\delta/N}$ and $\bar\theta$ the angle with respect 
to the $x$ axis. In this case, the 
$N\to \infty$ behavior of $NL_1$ can be shown to be quadratic in $\delta$, 
$NL_1=  2/(9 \sqrt{3}) \delta^2 + {\cal O}(\delta^3)$.


\subsection{Lower temperatures}
\label{low temperatures}

\begin{figure} [t]
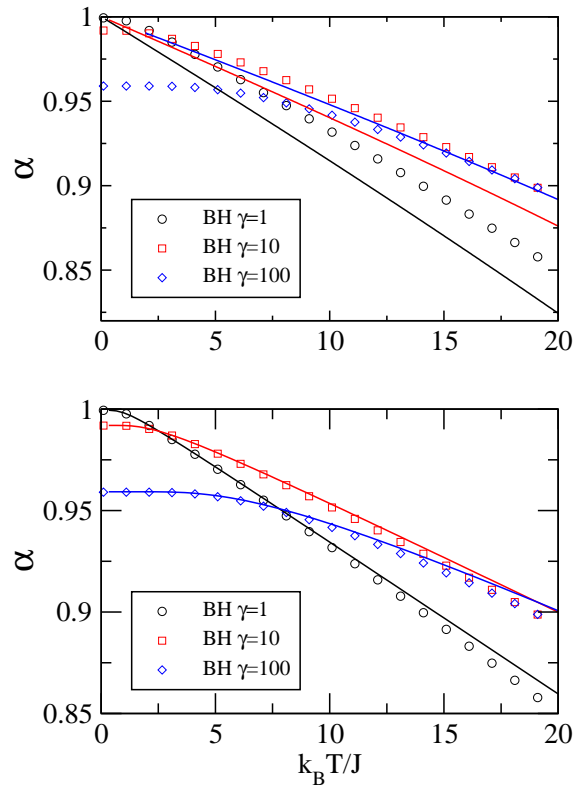
                         
\includegraphics[width=7.5cm, angle=0]{fig9b.eps}
\includegraphics[width=7.5cm, angle=0]{fig9a.eps}
\caption{
(Color online) Coherence factor of the system as a function of the temperature for 
several repulsive interactions, $\gamma=1,10$ and $100$, for $N=100$ atoms, 
focusing on the low temperature region where $k_B T\sim J$.  The symbols 
represent the BH results, and the solid curves are obtained from eq.~(\ref{eq:po7}) 
in the top panel and eq.~(\ref{eq:s8}) in the bottom panel.}
\label{fig:lowt}
\end{figure} 

At temperatures $k_B T \sim J$, the number of eigenvectors contributing 
to the canonical equilibrium thermal ensemble is small, and the GS 
theory breaks down. As seen in the top panel of Fig.~\ref{fig:lowt}, 
the results of exact BH predict a sizable loss of coherence at low 
temperatures, whereas the GS formula (\ref{eq:po7}) predicts full 
coherence as $T \rightarrow 0$.

Other approximations are available for this lower temperature 
regime~\cite{ober06,malo13}. For repulsive atom-atom interactions, 
the $h=1/N$ expansion discussed in Ref.~\cite{malo13} yields the 
following approximation of the coherence factor:
\beq
\alpha \ = \ {\cal A}-\frac{{\cal B}}{2} 
\left( \frac{N+1}{\tanh \beta F(N+1)} -\frac{1}{\tanh \beta F}-N\right)  
\label{eq:s8}
\eeq
with $\beta=1/k_B T$, $\;F = h \sqrt{\gamma+1-h}$, 
$\;{\cal B} = 2h \frac{-\gamma/2-1+h}{\sqrt{\gamma +1-h}}$, 
and $\;{\cal A} = 1+ h + \tfrac12 {\cal B}$. The bottom panel of 
Fig.~\ref{fig:lowt} shows that the incorporation of these 
$1/N$ effects produces very accurate results at $T\to 0$.

\section{Summary}
\label{sec:sum}

We have studied effects of relatively high temperatures on bosonic Josephson junctions, 
focusing on both attractive and repulsive atom-atom interactions in the ``Rabi-Josephson'' 
boundary regime where $|\gamma| \sim 1$. Our study proceeded by solving the $N$-particle 
model Hamiltonian (a two-site Bose Hubbard model) for moderately small $N$ (around $100$) 
and comparing the results to the predictions of the ``classical'' limit attained 
as $N\rightarrow \infty$.  

For temperatures much larger than the corresponding tunneling energy, the finite $N$ 
behavior (population imbalances and calculated Husimi distributions) is very close to 
the predictions of the classical limit, even for $N < 100$. Differences between the 
``exact'' finite-$N$ results and those of the classical limit appear to be proportional 
to $1/N$, with moderate correction coefficients.  

The current study may contribute to an understanding of the way that higher 
temperatures wash out quantum effects expected at $T=0$. Our main conclusion 
is that, when the temperature in a bosonic Josephson junction is so high, 
quantum effects will only be seen if the number of particles is rather small.

\begin{acknowledgments}
This work has been supported by FIS2011-24154 and 2009-SGR1289. 
B.~J.-D. is supported by the Ram\'on y Cajal program.  A.~G. acknowledges support by the Austrian Science Foundation (FWF) project
ViCoM (FWF No F41) and by the ANR-FWF project LODIQUAS (FWF No I830-N13). 
\end{acknowledgments}

\appendix

\section{Proof of the convergence to the classical limit}

In this appendix, we prove that the normalized Husimi distributions (\ref{normalized BH Husimis}) converge to the classical limit (\ref{normalized GS Husimi}), and then go on to prove the 
theorem of Ref.~\cite{GS09} concerning canonical averages of $k$-particle correlations.

The Bose-Hubbard Hamiltonian on the $N$-boson space can be written 
\begin{eqnarray*}
\lefteqn {H_N \ = \ 
-J (\hat{a}_1^{\dag} \hat{a}_2 + \hat{a}_2^{\dag} \hat{a}_1) 
+ \frac{U}{2}(\hat{a}_1^{\dag} \hat{a}_1^{\dag} \hat{a}_1 \hat{a}_1 
+ \hat{a}_2^{\dag} \hat{a}_2^{\dag} \hat{a}_2 \hat{a}_2 ) }
 \\
& = & 
- \frac{J}{N}
( \hat{a}_1^{\dag} \hat{a}_1^{\dag} \hat{a}_1\hat{a}_2  
+  \hat{a}_1^{\dag}\hat{a}_2^{\dag} \hat{a}_2 \hat{a}_2 
+ \hat{a}_2^{\dag} \hat{a}_1^{\dag} \hat{a}_1\hat{a}_1 
+ \hat{a}_2^{\dag} \hat{a}_2^{\dag} \hat{a}_2 \hat{a}_1 )
\\
&  & +\  \frac{U}{2}(\hat{a}_1^{\dag} \hat{a}_1^{\dag} \hat{a}_1 \hat{a}_1 
+ \hat{a}_2^{\dag} \hat{a}_2^{\dag} \hat{a}_2 \hat{a}_2 ) .
\end{eqnarray*}
In the preceding formula, operators like 
$\hat{a}_1^{\dag} \hat{a}_1^{\dag} \hat{a}_1\hat{a}_2$ are to be regarded as being 
restricted to the $N$-particle subspace, even though this 
is not indicated in the notation.  

With the dimensionless parameters $\delta$ and $\epsilon$ defined 
in Sec.~\ref{review of GS theory}, we have $-\tfrac{1}{k_BT}H_N  =  \tfrac{1}{N^2} W_N$,
where $W_N$ denotes the restriction of 
\begin{eqnarray*}
  W &=& 
\delta(\hat{a}_1^{\dag} \hat{a}_1^{\dag} \hat{a}_1\hat{a}_2  
+  \hat{a}_1^{\dag}\hat{a}_2^{\dag} \hat{a}_2 \hat{a}_2 
+ \hat{a}_2^{\dag} \hat{a}_1^{\dag} \hat{a}_1\hat{a}_1 
+ \hat{a}_2^{\dag} \hat{a}_2^{\dag} \hat{a}_2 \hat{a}_1 ) \\
&&-  \epsilon (\hat{a}_1^{\dag} \hat{a}_1^{\dag} \hat{a}_1 \hat{a}_1 
+ \hat{a}_2^{\dag} \hat{a}_2^{\dag} \hat{a}_2 \hat{a}_2 )
\end{eqnarray*}
to the $N$-particle component of the boson Fock space.
We are going to consider a limit where $N$ tends to infinity while 
$\delta$ and $\epsilon$ remain constant.  

For each point on the Bloch sphere with spherical coordinates 
$(\theta,\phi)$, let 
\[
     |\Psi_{\theta,\phi}\rangle \ = \ \big(\cos \tfrac{\theta}{2} \ a^{\dag}_1 
+ e^{ i \phi} \sin  \tfrac{\theta}{2}  \ a^{\dag}_2 \big) |\O \rangle
\]  
denote the corresponding mode, and recall that $|\Psi_{\theta,\phi}^N\rangle$ denotes   
the coherent state of $N$ bosons in that mode
(cf., formula (\ref{eq:cohs})).  Then, for any modes $|\chi_1\rangle, \ldots, |\chi_n\rangle$ and $|\chi'_1\rangle, \ldots, |\chi'_n\rangle$, 
\beqa
\lefteqn{ \big\langle \Psi_{\theta,\phi}^N   
\big| a^{\dagger}_{\chi_1}\cdots a^{\dagger}_{\chi_n} a_{\chi'_1}\cdots a_{\chi'_n} 
\Psi_{\theta,\phi}^N   \big\rangle }
 \nonumber \\
&=& \  N^{(n)}  \prod_{i=1}^n  \langle \chi'_i |  
\Psi_{\theta,\phi} \rangle\!\langle \Psi_{\theta,\phi} | \chi_i \rangle\ , 
\label{important}
\eeqa
where $N^{(n)} = \prod\limits_{m=0}^{n-1}(N-m)$.  
In particular, (\ref{important}) implies that 
\[
\big\langle \Psi_{\theta,\phi}^N   
\big| W_N \Psi_{\theta,\phi}^N   \big\rangle \ = \    
 N(N-1) (\delta x  - \tfrac12\epsilon z^2 -\tfrac12\epsilon )
\]
with 
\begin{eqnarray*}
x & = &  \sin\theta \cos\phi \\
z & = &  \cos \theta\ .
\end{eqnarray*}

 Let $\Xi$ denote a product of creators and annihilators  which, when normally ordered, becomes 
\[ 
   :\!\Xi\!:  \ = \  a^{\dagger}_{\chi_1}\cdots 
a^{\dagger}_{\chi_n} a_{\chi'_1}\cdots a_{\chi'_n}.
\]
Using the canonical commutation relations,  formula~(\ref{important}) 
implies that 
\beqa
\lefteqn{
 \lim\limits_{N \rightarrow \infty} 
\frac{1}{N^n} \big\langle \Psi_{\theta,\phi}^N   
\big| \Xi \Psi_{\theta,\phi}^N   \big\rangle 
} \nonumber \\
& = & 
 \lim\limits_{N \rightarrow \infty} 
\frac{1}{N^n} \big\langle \Psi_{\theta,\phi}^N   
\big| :\!\Xi\!: \Psi_{\theta,\phi}^N   \big\rangle  \nonumber \\
& = & 
\prod_{i=1}^n  \langle \chi'_i |  \Psi_{\theta,\phi}^N \rangle\!
\langle  \Psi_{\theta,\phi}^N | \chi_i \rangle\ .
\label{coherent state expectation} 
\eeqa
Thus
\begin{eqnarray*}
 \lefteqn{  \lim\limits_{N \rightarrow \infty} \frac{1}{N^{2j}} 
\big\langle \Psi_{\theta,\phi}^N   \big| W_N^j \ \Psi_{\theta,\phi}^N 
\big\rangle } \\
& = & 
      \lim\limits_{N \rightarrow \infty} \frac{1}{N^{2j}} 
\big\langle \Psi_{\theta,\phi}^N   \big| :\!W_N^j\!: \Psi_{\theta,\phi}^N 
\big\rangle 
\ = \    \left( \delta x  
- \tfrac12\epsilon z  -\tfrac12\epsilon \right)^j
\\
\end{eqnarray*}
and therefore 
\begin{eqnarray}
\lefteqn{  \lim_{N\rightarrow\infty}  
\big\langle \Psi_{\theta,\phi}^N  | e^{-H_N/k_BT}
\Psi_{\theta,\phi}^N  \big\rangle } \nonumber\\
  & = & 
  \lim_{N\rightarrow\infty}  
\big\langle \Psi_{\theta,\phi}^N  | \exp(\tfrac{1}{N^2} W_N) 
\Psi_{\theta,\phi}^N  \big\rangle \nonumber\\
    & = &  
\sum_{j=0}^{\infty}  \frac{1}{j!} \lim_{N\rightarrow\infty} \frac{1}{N^{2j}} \langle 
\Psi_{\theta,\phi}^N  |W_N^j \Psi_{\theta,\phi}^N  \rangle \nonumber  \\
& = &  
\exp \left(\delta x  - \tfrac{\epsilon}{2} z^2 -\tfrac{\epsilon}{2} \right) \,.
\label{limit of canonical density}
\end{eqnarray}

The Husimi distribution $P_{N,T}(\theta,\phi)$ defined in (\ref{Husimi of canonical}) is proportional to 
$\big\langle \Psi_{\theta,\phi}^N  | e^{-H_N/k_BT}\Psi_{\theta,\phi}^N  \big\rangle$.  Its normalized version $\tilde{P}_{N,T}$ defined in formula (\ref{normalized BH Husimis}) integrates to $1$.  It follows from (\ref{limit of canonical density}) that $\tilde{P}_{N,T}$ converges to $\tilde{Q}_{\delta,\epsilon}$ of formula (\ref{normalized GS Husimi}) in the limit considered.

We now proceed to derive the theorem of Gottlieb and Schumm concerning canonical averages of $k$-particle correlations.  This is the result that is paraphrased near the beginning of Section~\ref{review of GS theory}, but we prove it here in a slightly more general form, equivalent Theorem~1 of \cite{GS09} for the case of $M=2$ modes.   
That is, we prove the following:
\begin{theorem}
Let 
\[ 
   X  \ = \  a^{\dagger}_{\chi_1}\cdots 
a^{\dagger}_{\chi_k} a_{\chi'_1}\cdots a_{\chi'_k}
\]
be a simple $k$-body operator.  Then, in the limit where $N$ tends to infinity while 
$\delta$ and $\epsilon$ remain constant, 
\begin{eqnarray*}
\lefteqn{ \lim_{N\rightarrow\infty}  \frac{1}{N^k} 
\big\langle X \big\rangle_{N,T} } \\
& = &  
 \frac{1}{{\cal N} } \int_S  dS\ \prod_{i=1}^k  \langle \chi'_i |  
\Psi_{\theta,\phi} \rangle\! \langle \Psi_{\theta,\phi} | \chi_i \rangle
 \ e^{\delta x - \tfrac{\epsilon}{2} z^2 }\ ,
\end{eqnarray*}
where ${\cal N}$ is the normalizing constant (\ref{normalizing const}).
\end{theorem}

\noindent{\bf  Proof:} \qquad 
As in formula (\ref{coherent state expectation}), we have 
\begin{eqnarray*}
\lefteqn{    \lim\limits_{N \rightarrow \infty} \frac{1}{N^{2j+k}} 
\big\langle \Psi_{\theta,\phi}^N   \big| W_N^j X\ \Psi_{\theta,\phi}^N 
\big\rangle } 
 \\
     & = &  
      \lim\limits_{N \rightarrow \infty} \frac{1}{N^{2j+k}} 
\big\langle \Psi_{\theta,\phi}^N   \big| :\!W_N^j X\!: \Psi_{\theta,\phi}^N 
\big\rangle 
\\
     & = &  
     \prod_{i=1}^k  \langle \chi'_i |  \Psi_{\theta,\phi}^N \rangle\!
\langle  
\Psi_{\theta,\phi}^N | \chi_i \rangle \left( \delta x 
- \tfrac12\epsilon z^2 -\tfrac12\epsilon \right)^j. 
\end{eqnarray*}
Using the preceding formula and the completeness relation (\ref{completeness}), 
 i.e., the fact that 
\[
      \frac{N+1}{4\pi} \int_S dS  
\big|  \Psi_{\theta,\phi}^N  \big\rangle\!\big\langle \Psi_{\theta,\phi}^N \big|
\]
is the identity operator on the $N$-boson Hilbert space, we get  
\begin{eqnarray*}
\lefteqn{
\lim_{N\rightarrow\infty} 
\frac{1}{N^{2j+k+1}}\tr ( W_N^jX  ) } 
 \\
& = & 
 \frac{1}{4\pi}  \lim_{N\rightarrow\infty} \frac{N+1}{N^{2j+k+1}} 
\tr \left[  W_N^jX  \int_S dS \big|  
\Psi_{\theta,\phi}^N  \big\rangle\!\big\langle  \Psi_{\theta,\phi}^N   \big| \right]  
 \\
& = & 
 \frac{1}{4\pi} \lim_{N\rightarrow\infty} \frac{1}{N^{2j+k}} \int_S dS \big\langle  
\Psi_{\theta,\phi}^N   
\big|W_N^jX   \ \Psi_{\theta,\phi}^N \big\rangle  
 \\
& = & 
 \frac{1}{4\pi}  \int_S dS \prod_{i=1}^k  \langle \chi'_i |  \Psi_{\theta,\phi}^N \rangle\!
\langle  \Psi_{\theta,\phi}^N | \chi_i \rangle 
\left( \delta x 
- \tfrac12\epsilon\ z^2 -\tfrac12\epsilon \right)^j
\end{eqnarray*}
and therefore
\begin{eqnarray*}
 \lefteqn{  \lim\limits_{N\rightarrow\infty}  \frac{1}{N^{k+1}} \tr \big[ e^{-H_N/k_BT} X  \big]  } \\
 & = &   \lim_{N\rightarrow\infty} \frac{1}{N^{k+1}} \tr \Big[ \exp(\tfrac{1}{N^2} W_N) X  \Big]  
   \\
   & = & 
    \sum_{j=0}^{\infty}  \frac{1}{j!}  \lim_{N\rightarrow\infty}  
\frac{1}{N^{2j+k+1}}\tr ( W_N^jX  )
\\
   & = & 
   \frac{e^{-\tfrac{\epsilon}{2}} }{4\pi} 
\int_S   dS\  \prod_{i=1}^k  \langle \chi'_i |  
\Psi_{\theta,\phi} \rangle\! \langle \Psi_{\theta,\phi} | 
\chi_i \rangle \ e^{\delta x  - \tfrac{\epsilon}{2} z^2 }.
\end{eqnarray*}  
In particular, 
\[
  \lim\limits_{N\rightarrow\infty}  \frac{1}{N} \tr \big[ e^{-H_N/k_BT}  \big]   
  \ = \    \frac{ e^{ -\tfrac{\epsilon}{2} } }{4\pi} 
\int_S   dS\  e^{\delta x - \tfrac{\epsilon}{2} z^2 }.
\]
The two preceding limits imply that  
\begin{eqnarray*}
\lefteqn{ \lim_{N\rightarrow\infty}  \frac{1}{N^k} \big\langle a^{\dagger}_{\chi_1}\cdots 
a^{\dagger}_{\chi_k} a_{\chi'_1}\cdots a_{\chi'_k} \big\rangle_{N,T} }\\
& = & \lim_{N\rightarrow\infty}  \frac{1}{N^k} 
\frac{\tr \big[ e^{-H_N/k_BT} X  \big]}{\tr  \big[ e^{-H_N/k_BT}\big] } \\
& = & 
\frac{ \lim\limits_{N\rightarrow\infty}  \tfrac{1}{N^{k+1}}   \tr \big[ e^{-H_N/k_BT} X  \big]}{ \lim\limits_{N\rightarrow\infty}  \tfrac{1}{N}  \tr \big[ e^{-H_N/k_BT}\big] } \\
& = &  
 \frac{1}{{\cal N} } \int_S dS\  \prod_{i=1}^k  \langle \chi'_i |  
\Psi_{\theta,\phi} \rangle\! \langle \Psi_{\theta,\phi} | \chi_i \rangle
 e^{\delta x  - \tfrac{\epsilon}{2} z^2 } .
\end{eqnarray*}


\vfill





\begin{thebibliography}{99}


\bibitem{Mil97} G.J. Milburn, J. Corney, 
E. M. Wright, and D. F. Walls, 
 Phys. Rev. A {\bf 55}, 4318 (1997).

\bibitem{Smerzi97}
A. Smerzi, S. Fantoni, S. Giovanazzi, and S. R. Shenoy, 
Phys. Rev. Lett. {\bf 79}, 4950 (1997).

\bibitem{cirac98} 
J. I. Cirac, M. Lewenstein, K. Molmer, and 
P. Zoller, Phys. Rev. A {\bf 57}, 1208 (1998).

\bibitem{jav99} J. Javanainen M. Y. Ivanov 
Phys. Rev. A {\bf 60}, 2351 (1999).

\bibitem{raghavan99}
S. Raghavan, A. Smerzi, S. Fantoni, and S. R. Shenoy
Phys. Rev. A {\bf 59}, 620 (1999).

\bibitem{LE01} 
A. J. Leggett, 
Rev. Mod. Phys. {\bf 73}, 307 (2001).

\bibitem{jame05} 
M. J\"a\"askel\"ainen, and P. Meystre, 
Phys. Rev. A {\bf 71}, 043603 (2005); 
Phys. Rev. A {\bf 73}, 013602 (2006).

\bibitem{mueller06} E. J. Mueller, T-L. Ho, M. Ueda, and 
G. Baym, Phys. Rev. A {\bf 74}, 033612 (2006).

\bibitem{ours10} B. Juli\'a-D\'{\i}az, D. Dagnino, 
M. Lewenstein, J. Martorell, and A. Polls, 
Phys. Rev. A {\bf  81}, 023615 (2010).

\bibitem{mom10}
C. Ottaviani, V. Ahufinger, R. Corbal\'an, J. Mompart, 
Phys. Rev. A {\bf 81}, 043621 (2010).


\bibitem{grae08}
M. P. Strzys, E. M. Graefe and H. J. Korsch, 
New J. Phys. {\bf 10}, 013024 (2008).

\bibitem{vardi13}
C. Khripkov, D. Cohen, A. Vardi, Phys. Rev. E {\bf 87}, 012910 (2013).

\bibitem{albiez05}
M. Albiez, R. Gati, J. F\"olling, S. Hunsmann, M. Cristiani, and 
M.K. Oberthaler, Phys. Rev. Lett. {\bf95}, 010402 (2005).


\bibitem{zib10} 
T. Zibold, E. Nicklas, C. Gross, 
and M. K. Oberthaler, 
Phys. Rev. Lett. {\bf 105}, 204101 (2010).

\bibitem{gross12} 
C. Gross, J. Phys. B: At. Mol. Opt. Phys. {\bf 45}, 103001 (2012). 

\bibitem{lipkin} H.J. Lipkin, N. Meshkov, 
and A.J. Glick, Nucl. Phys. {\bf 62}, 188 (1965).

\bibitem{esteve08} 
J. Esteve, C. Gross, A. Weller, S. Giovanazzi, 
and M. K. Oberthaler, Nature {\bf 455}, 1216, (2008).

\bibitem{gross10} C. Gross, T. Zibold, E. Nicklas, 
J. Est\`eve, and M. K. Oberthaler,  
Nature {\bf 464}, 1165 (2010).

\bibitem{zibphd} T. Zibold, PhD-Thesis: ``Classical Bifurcation and 
Entanglement Generation in an Internal Bosonic Josephson Junction'', 
U. Heidelberg (2012).

\bibitem{ours11} B. Juli\'a-D\'iaz, J. Martorell, A. Polls, 
Phys. Rev. A {\bf 81}, 063625 (2010). 

\bibitem{kita} 
M. Kitagawa, and M. Ueda, 
Phys. Rev. A {\bf 47}, 5138 (1993).

\bibitem{vardi} J. R. Anglin, and, A. Vardi, 
Phys. Rev. A {\bf 64}, 013605 (2001); A. Vardi, 
and J. R. Anglin, Phys. Rev. Lett. {\bf 86}, 568 (2001).

\bibitem{PS01} L. Pitaevskii and S. Stringari, 
Phys. Rev. Lett. {\bf 87}, 180402 (2001). 

\bibitem{GO07} R. Gati and M.K. Oberthaler, 
J. Phys. B: At. Mol. Opt. Phys. {\bf 40},  R61 (2007).

\bibitem{GS09} A.D. Gottlieb and T. Schumm,  
Phys. Rev. A {\bf 79}, 063601 (2009)
 

\bibitem{ober06} 
R. Gati, B. Hemmerling, J. F\"olling, M. Albiez, and M. K. Oberthaler
Phys. Rev. Lett. {\bf 96}, 130404 (2006).

\bibitem{malo13} B. Juli\'a-D\'iaz, J. Martorell, and A. Polls, 
{\it Spontaneous symmetry breaking, self-trapping and Josephson 
oscillations}, Progress in optical science and photonics, ed. B. Malomed, 
Springer (2013).

\bibitem{husiref} F. T. Arecchi, E. Courtens, R. Gilmore, and H. Thomas.  
Phys. Rev. A {\bf 6}, 2211 (1972).

\bibitem{lee84}
C.T. Lee,
Phys. Rev. A 30 3308 - 3310

\bibitem{mahmud05} 
K. W. Mahmud, H. Perry, and W. P. Reinhardt, Phys. Rev. A {\bf 71}, 023615 (2005).

\bibitem{oursober} 
B. Juli\'a-D\'iaz, T. Zibold, M. K. Oberthaler, M. Mele-Messeguer, 
J. Martorell, A. Polls, Phys. Rev. A {\bf 86}, 023615 (2012).

\bibitem{sore01} A. S. S\o rensen, L. M. Duan, J. I. Cirac, and 
P. Zoller, Nature {\bf 409}, 63 (2001).

\bibitem{noise_thermometer06} 
R. Gati, J. Esteve, Hemmerling, T. B. Ottenstein, J. Appmeier, 
A. Weller and. M K Oberthaler,
New J. Phys. {\bf 8}, 189 (2006).

\bibitem{ST08} V. S. Shchesnovich, and 
M. Trippenbach, 
Phys. Rev. A {\bf 78}, 023611, (2008).




\end{thebibliography}
\end{document}